\documentclass[]{spie}           % use for US letter paper
\usepackage[dvips]{graphicx}

\DeclareTextSymbolDefault{\micro}{TS1}
\DeclareTextSymbol{\micro}{TS1}{181} % micro sign
\DeclareFontEncoding{TS1}{}{}
\DeclareFontSubstitution{TS1}{cmr}{m}{n}

\title{A catalog of reference stars \\
for long baseline stellar interferometry\footnote{The catalog is available
from the authors.}} 

\author{Pascal Bord\'e\supit{a}, Vincent Coud\'e du Foresto\supit{a},
Gilles Chagnon\supit{a}, and Guy Perrin\supit{a}
\skiplinehalf
\supit{a}Laboratoire d'\'Etudes Spatiales et d'Instrumentation en
Astrophysique, \\ Observatoire de Paris, Meudon, France
}

\authorinfo{Further author information: (Send correspondence to P.B.) \\
P.B.: E-mail: Pascal.Borde@obspm.fr \\
Copyright 2002 Society of Photo-Optical Instrumentation Engineers. \\
This paper will be published in the proceedings of the conference
\emph{Interferometry for Optical Astronomy II}, part of SPIE's Astronomical
Telescopes \& Instrumentation, 22--28 August 2002, Waikoloa, HI and is made
available as an electronic preprint with permission of SPIE. One print or
electronic copy may be made for personal use only. Systematic or multiple
reproduction, distribution to multiple locations via electronic or other means,
duplication of any material in this paper for a fee or for commercial purposes,
or modification of the content of the paper are prohibited.
}
 
\begin{document}
  \maketitle 

%_____________________________________________________________________ Abstract
%
\begin{abstract}
The calibration process of long baseline stellar interferometers requires the
use of reference stars with accurately determined angular diameters. We present
a catalog of 374 carefully chosen stars among the all-sky network of infrared
sources provided by Ref.~\citenum{Cohen99}. The catalog benefits from a very
good sky coverage and a median formal error on the angular diameters of only
1.2\%. Besides, its groups together in a homogeneous handy set stellar
coordinates, uniform and limb-darkened angular diameters, photometric
measurements, and other parameters relevant to optical interferometry. In this
paper, we describe the selection criteria applied to qualify stars as reference
sources. Then, we discuss the catalog's statistical properties such as the
sky coverage or the distributions of magnitudes and angular diameters. We study
the number of available reference stars as a function of the baseline and the
precision needed on the visibility measurements. Finally, we compare the angular
diameters predicted in Ref.~\citenum{Cohen99} with existing determinations in
the literature, and find a very good agreement.
\end{abstract}

\keywords{catalogs, stars: fundamental parameters,
instrumentation: interferometers, techniques: interferometric}

%_______________________________________________________________ 1 Introduction
%
\section{INTRODUCTION} \label{sec:intro}
Astronomical optical interferometers need to be calibrated not only against
long term drifts but also against short term effects due to the atmospheric
turbulence. The usually adopted solution consists in interleaving observations
of scientific targets and reference sources. A reference source is an
astronomical source for which the theoretical fringe contrast, or visibility,
can be predicted with a high accuracy. The visibility of the scientific target
can then be deduced from the equation
\begin{equation} \label{eq:cal}
V_\mathrm{target} = \frac{\mu_\mathrm{target}}{\mu_{\mathrm{ref}}}
V_\mathrm{ref},
\end{equation}
where $\mu$ denotes a measured fringe contrast and $V$ a visibility.
The reference sources that have the most simple model are non-resolved or
almost non-resolved single stars with compact atmospheres, and will be called
reference stars or calibrators in the following. They can be correctly
described by a uniform disk (UD) model whose visibility is
\begin{equation} \label{eq:UD}
V_\mathrm{UD} = \frac{2 J_1(x)}{x}
\quad \mbox{with} \quad x = \pi \sigma_\mathrm{eff} B \, \theta_\mathrm{UD},
\end{equation}
where $\sigma_\mathrm{eff}$ is the effective wavenumber (see
Sect.~\ref{sub:ew_sf}), $B$ the interferometric baseline projected on the sky,
and $\theta_\mathrm{UD}$ the stellar angular diameter. As many instrumental
effects depend on the direction aimed at the sky, it is preferable that the
reference star be close to the target. Hence arises the need for a grid of
such stars with a good sky coverage. In this paper, we describe the catalog
of reference stars that was made up for that purpose by the
FLUOR\cite{Foresto98} team (more details will be available in a forthcoming
paper\cite{Borde02}).

In Sect.~\ref{sec:sel}, we explain the selection process of our reference
stars. Section~\ref{sec:cont} describes the catalog's content and
Sect.~\ref{sec:stat} its statistical properties. Finally, we compare the
angular diameters of our reference stars to other existing determinations in
Sect.~\ref{sec:comp}.
%
%_______________________________________________ 2 Selection of reference stars
%
\section{SELECTION OF REFERENCE STARS} \label{sec:sel}
As explained in the introduction, reference stars have to be non-variable
single stars with compact atmospheres and accurately known angular diameters.
In order to build there all-sky network of absolutely calibrated stellar
spectra, Martin Cohen and collaborators\cite{Cohen99} have used criteria that
match quite well are requirements. Moreover, they have derived angular
diameters with formal errors by fitting Kurucz's atmosphere models to stellar
spectra of some prototype stars. By making the fundamental assumption that every
K0--M0 giant has a spectrum identical to its prototype, they have extended
their collection of spectra to 422 well chosen stars by rescaling them in flux
thanks to photometric measurements. Angular diameters are then derived using the
scaling factor. In the following, this method will referred to as the
spectro-photometric method (SPM).

We have taken advantage of this existing network by extracting a subset of
reference stars suitable for the calibration of stellar interferometers. Our
extra requirements are essentially the absence of significantly variable
($>0.01$~mag) and close binary stars that would both necessitate a model more
elaborate than Eq.~(\ref{eq:UD}). The initial network is then cross-checked
with the Simbad database\footnote{http://simbad.u-strasbg.fr/Simbad}, the
Batten catalog of spectroscopic binaries\cite{Batten89}, and the catalog of
visual double stars observed by Hipparcos\cite{Dommanget00}. We choose to
discard all double stars with separations less than 4$''$, and to avoid
pointing confusion, we keep double stars with separations between 4$''$ and
30$''$ only when the companion is five magnitudes fainter than the primary.
Companions' magnitudes and separations are notified in the comments. As a
result of this more stringent selection, our catalog is left wih 374 entries.
%
%__________________________________________________________ 3 Catalog's content
%
\section{CATALOG'S CONTENT} \label{sec:cont}
\begin{table}[h]
\caption{Summary of the catalog's content.} 
\label{tab:cont}
\begin{center}       
\begin{tabular}{|l|l|}
\hline
\rule[-1ex]{0pt}{3.5ex}  Star identification & HD and HR numbers,  Bayer/Flamsteed name \\
\hline
\rule[-1ex]{0pt}{3.5ex}  Coordinates         & Right ascension, declination, proper motions, parallax \\
\hline
\rule[-1ex]{0pt}{3.5ex}  Physical properties & Spectral type, effective temperature, surface gravity, \\
\rule[-1ex]{0pt}{3.5ex}                      & linear limb-darkening coefficients in the J, H, and K bands \\
\hline
\rule[-1ex]{0pt}{3.5ex}  Angular diameters   & LD diameters, UD diameters in the J, H, and K bands \\
\hline
\rule[-1ex]{0pt}{3.5ex}  Cross-properties    & Effective wavenumber and shape factor in the K' band \\
\hline
\rule[-1ex]{0pt}{3.5ex}  Photometry          & B, V, J, H, K, L, M, and N Johnson's magnitudes \\
\hline
\rule[-1ex]{0pt}{3.5ex}  Comments            & Simbad classification, companions' magnitudes and separations if any \\
\hline 
\end{tabular}
\end{center}
\end{table}
%
%______________________________________________________ 3.1 Star identification
%
\subsection{Star identification}
The catalog is meant to group together all useful information in the context
of long baseline stellar interferometry (LBSI). The Henry Draper (HD) number
has been chosen as the main identifier in the catalog (the Bright Star
Catalog number, denoted HR, is also provided for convenience). As the
knowledge of these stars is likely to be improved in the future, it is very
important to keep track of the calibrator(s) used for a given scientific
observation, so that any data could be reduced again if necessary.
Additionally, it makes easier the search for observations that used the same
calibrator(s) and that are thus correlated\cite{Perrin02}.
Identifiers (HD, HR, and Bayer or Flamsteed name) are followed by the
stellar coordinates, some physical properties, angular diameters in different
bands, some cross-properties of the star and FLUOR, the photometry, and some
comments (Table~\ref{tab:cont}).
%
%________________________________________________________ 3.2 Angular diameters
%
\subsection{Angular diameters} \label{sub:diam}
Limb-darkened angular diameters have been computed in Ref.~\citenum{Cohen99}
for every star. This diameter corresponds to the physical diameter of the star,
\textit{i.e.} the one that appears in the Stefan-Boltzmann law
\begin{equation} \label{eq:stef_boltz}
F_\mathrm{bol} = \frac{1}{4} \, \theta_\mathrm{LD}^2 \, \sigma_\mathrm{S} \,
T_\mathrm{eff}^4,
\end{equation}
where $F_\mathrm{bol}$ is the bolometric flux (W/m$^2$) emitted by the star and
$\sigma_\mathrm{S}$ denotes Stefan-Boltzmann constant. As such,
$\theta_\mathrm{LD}$ is independent of the observational wavelength.
This diameter can be converted into the UD angular diameter of
Eq.~(\ref{eq:UD}), usually used by interferometrists. However, the latter
depends on the wavelength, so a spectral band has to be specified. The following
formula\cite{Hanbury74} provides an efficient way to perform the conversion
using linear limb-darkening coefficients $u_\lambda$:
\begin{equation} \label{eq:ud}
\frac{\theta_\mathrm{LD}}{\theta_\mathrm{UD}} = \sqrt{\frac{1-u_\lambda/3}
{1-7u_\lambda/15}}.
\end{equation}
For every star, we interpolate $u_\lambda$ into the tables computed in
Ref.~\citenum{Claret95} using the effective temperature $T_\mathrm{eff}$ and
the surface gravity $\log(g)$ derived from the spectral
type\cite{Jager87,Straizys81}. Then, Eq.~\ref{eq:ud} yields UD angular
diameters in the J, H, and K bands. As the conversion process introduces an
additional although very small error, the catalog states the new uncertainty
for every UD diameter.
%
%_______________________________________________________________ 3.3 Photometry
%
\subsection{Photometry}
For every star, the catalog features the B and V magnitudes drawn from the
Simbad database, and the J to N infrared magnitudes taken from
Ref.~\citenum{Cohen99}, or estimated from the spectral type using the tables in
Refs.~\citenum{Cox00} and \citenum{Johnson66}. A boolean flag indicates whether
the quoted value is a measurement or not.
%
%____________________________________ 3.4 Effective wavenumber and shape factor
%
\subsection{Effective wavenumber and shape factor} \label{sub:ew_sf}
The effective wavenumber and the shape factor are cross-properties of the
star's spectrum and of the instrument. In the case of FLUOR, observations
are carried out in the K' band (2.0--2.3~{\micro}m) and these quantities have
been computed in this band only. The effective wavenumber is the wavenumber at
which the monochromatic visibility defined by Eq.~(\ref{eq:UD}) is equal to the
measured wide-band visibility. If $\mathcal{S}$ denotes the star's spectrum
multiplied by the filter's transmission profile, the effective wavenumber is
\begin{equation}
\sigma_{\mathrm{eff}} \equiv \frac{\int_0^{\infty} \sigma \,
\mathcal{S}^2(\sigma) \mathrm{d}\sigma} {\int_0^{\infty} \mathcal{S}^2(\sigma)
\, \mathrm{d}\sigma}.
\end{equation}
As explained in Ref.~\citenum{Foresto97}, the wide-band fringe contrast measured
by FLUOR is weighted by the squared stellar spectrum:
\begin{equation}
\overline{\mu^2} = \frac{1}{S\!F} \int_0^{\infty} \mu^2(\sigma)
\mathcal{S}^2(\sigma) \, \mathrm{d}\sigma
\quad \mathrm{with} \quad
S\!F \equiv \int_0^{\infty} \mathcal{S}^2(\sigma) \, \mathrm{d}\sigma.
\end{equation}
The shape factor $S\!F$ allows for a correct calibration when the spectral
types of the target and its reference stars are different. Effective wavenumbers
and shape factors should be mostly considered as relative information between
stars of different spectral types. Their typical values are respectively
4685~cm$^{-1}$ and 13.19~{\micro}m, and vary very little from one spectral type
to another.
%
%_________________________________________________________________ 3.5 Comments
%
\subsection{Comments}
This field is used to provide additional information about the source: the
object type\cite{Ochsenbein92} as it is given by Simbad, the separations and
the magnitudes of the companions when the source is a double or a multiple
star.
%
%___________________________________________ 4 Catalog's statistical properties
%
\section{CATALOG'S STATISTICAL PROPERTIES} \label{sec:stat}
%
%________________________________________________________ 4.1 Global statistics
%
\subsection{Global statistics}
\begin{figure}[htbp]
\begin{center}
\begin{tabular}{c}
\includegraphics[width=8cm]{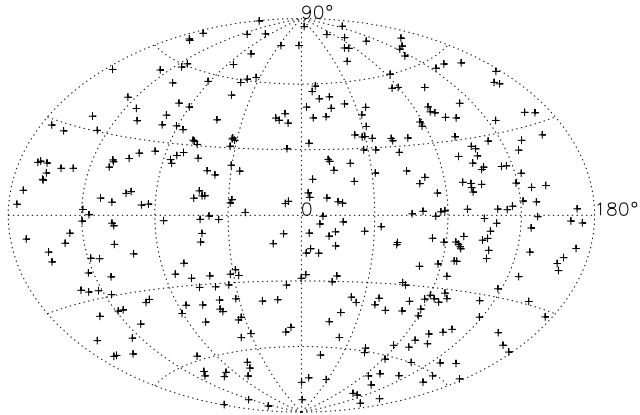}
\end{tabular}
\end{center}
\caption{ \label{fig:sky} Catalog's sky coverage in the Hammer-Aitoff
equal-area projection.}
\end{figure}

\begin{figure}[htbp]
\begin{center}
\begin{tabular}{c}
\includegraphics[width=16cm]{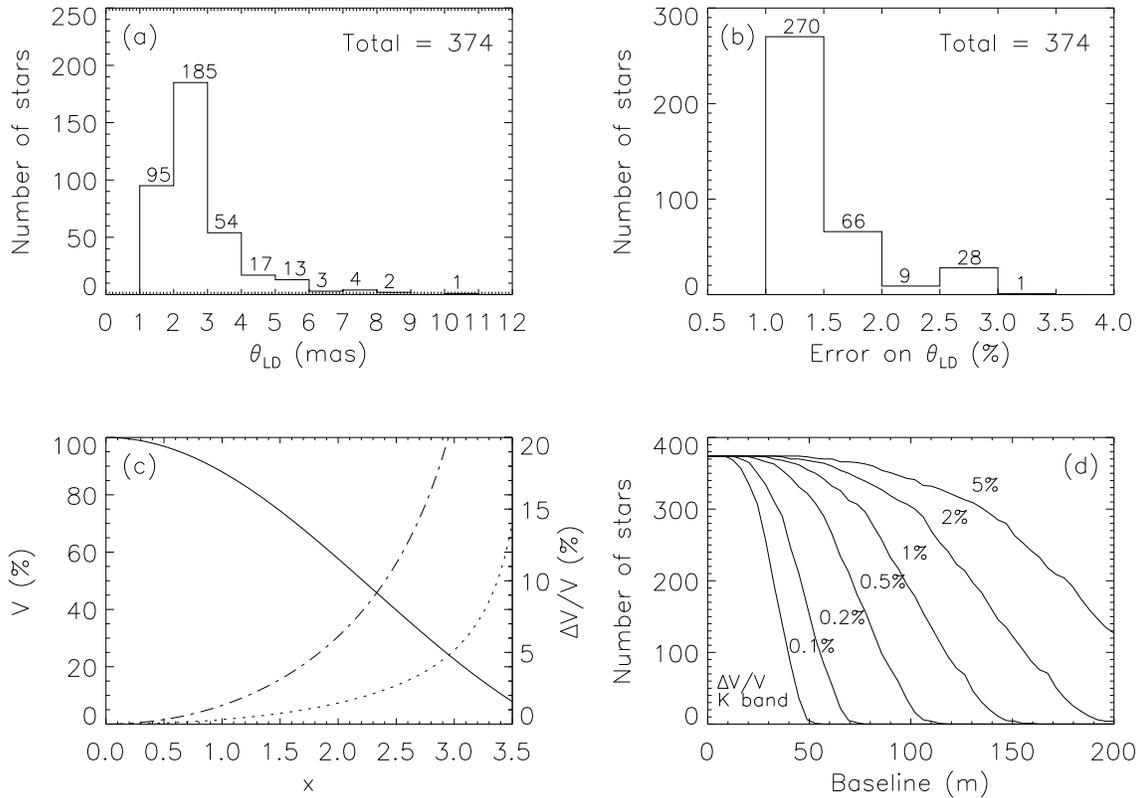}
\end{tabular}
\end{center}
\caption{ \label{fig:stat} (a) Histograms of limb-darkened (LD) diameters and
(b) their associate errors. (c) Visibility of a uniform disk (UD) and relative
error on the visibility due to an error on the UD diameter of respectively 1.2\%
(dotted line) and 5\% (dash-dotted line). (d) Catalog's effective size in the
K band: number of stars whose formal errors on the UD diameter in the K band
make them suitable reference stars at the labeled relative precision on the
visibility.}
\end{figure}

A major feature of our catalog is its excellent sky coverage
(Fig.~\ref{fig:sky}): whatever the point on the sky, its distance to the
closest reference star is less than 16.4$^\circ$ and the median distance is
5.2$^\circ$. 
Most stars (91\%) are class~III giants with a spectral type K (82\%) or M0
(18\%). Also most stars (72\%) have a visual magnitude between 4 and 6,
and	almost all of them (95\%) between 3 and 7, with a median value of 5.0. As
for K magnitude, most stars (95\%) lie in the interval K=0--3 with a median
value of 1.8. Limb-darkened angular diameters range from 1 to 10 mas
(Fig.~\ref{fig:stat}a) with a median value of 2.3 mas. The median error on the
diameter is only 1.2\% (Fig.~\ref{fig:stat}b), which brings a significant gain
(Fig.~\ref{fig:stat}c) compared to the classical 5--10\% (\textit{e.g.}
Ref.~\citenum{Boden00}) encountered when little is known about the calibrator.
%
%_________________________________________________ 4.2 Catalog's effective size
%
\subsection{Catalog's effective size}
In the framework of a UD model, the relative error on the visibility reads
\begin{equation} \label{eq:err}
\frac{\Delta V_\mathrm{UD}}{V_\mathrm{UD}} = x \frac{J_2(x)}{J_1(x)}
\frac{\Delta \theta_\mathrm{UD}}{\theta_\mathrm{UD}}
\quad \mbox{with} \quad x = \pi \sigma_\mathrm{eff} B \, \theta_\mathrm{UD},
\end{equation}
assuming negligeable errors on the effective wavenumber and on the
interferometric baseline. Figure~\ref{fig:stat}c represents the visibility and
the error on the visibility as a function of the reduced variable $x$.
Equation~(\ref{eq:err}) implies that the catalog's reference stars have not all
diameter estimates accurate enough to allow a given precision on the visibility,
whatever the wavenumber and the baseline. For example, the number of stars whose
error on the angular diameter is small enough to be used for a given accuracy on
the visibility in the K band and at a given baseline is given in
Table~\ref{tab:effsz}, as well as displayed on Fig.~\ref{fig:stat}d.
\begin{table}[h]
\caption{Catalog's effective size in the K band: number of stars whose error
on the angular diameter is small enough to be used for a given accuracy on the
visibility.} 
\label{tab:effsz}
\begin{center}       
\begin{tabular}{|r|rrrr|}
\hline
\rule[-1ex]{0pt}{3.5ex}           & & $\Delta V/V$ & &     \\
\rule[-1ex]{0pt}{3.5ex}  Baseline & $\le0.5\%$ & $\le1\%$ & $\le2\%$ & $\le5\%$ \\
\hline
\rule[-1ex]{0pt}{3.5ex}      50 m & 316 & 354 & 366 & 372 \\
\rule[-1ex]{0pt}{3.5ex}     100 m &  24 & 186 & 305 & 341 \\
\rule[-1ex]{0pt}{3.5ex}     150 m &   0 &   4 & 126 & 266 \\
\rule[-1ex]{0pt}{3.5ex}     200 m &   0 &   0 &   4 & 127 \\
\hline
\end{tabular}
\end{center}
\end{table}
%
%______________________________ 5 Comparison with other diameter determinations
%
\section{COMPARISON WITH OTHER DIAMETER DETERMINATIONS} \label{sec:comp}
We have searched the literature by the way of CHARM\footnote{Catalog of High
Angular Resolution Measurements} catalog\cite{Richichi02} for other angular
diameter determinations of the stars in our catalog. Angular diameters can
either be estimated by photometric means or directly measured by LBSI or during
a lunar occultation. In the following sections, we will examine two photometric
methods and direct measurements performed by two interferometers.
%
%______________________________________________________ 5.1 Photometric methods
%
\subsection{Photometric methods}
\begin{figure}[htbp]
\begin{center}
\begin{tabular}{lll}
\includegraphics[width=16cm]{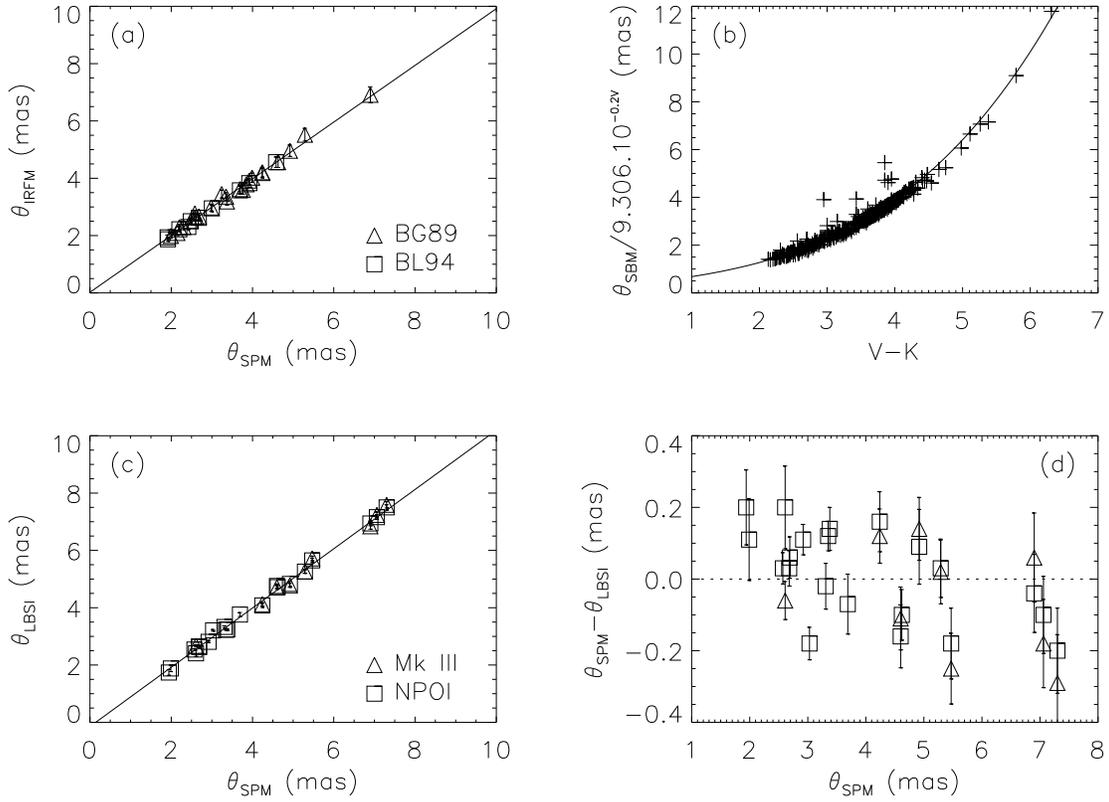}
\end{tabular}
\end{center}
\caption{ \label{fig:comp} (a) LD angular diameters determined by the
infrared flux method (IRFM) vs. those determined by the spectro-photometric
method (SPM). The solid line represents the least-square linear fit to the
data. (b) Comparison between SPM angular diameters (crosses) and those (solid
line) predicted by the surface brightness method (SBM) as a function of V$-$K.
(c) LD angular diameters determined by long baseline stellar interferometry
(LBSI) vs. those determined by the SPM. The solid line represents the
least-square linear fit to the data. (d) Difference between SPM and LBSI
diameters vs. SPM diameters.}
\end{figure}

The angular LD diameters of 29 stars belonging to our catalog have been computed
with the infrared flux method (IRFM) and are reported either in
Ref.~\citenum{Bell89} or \citenum{Blackwell94}. As can be seen on
Fig.~\ref{fig:comp}a, the agreement with the spectro-photometric
method\cite{Cohen99} (SPM) is excellent: a linear least-square fit to the data
yields $\theta_\mathrm{IRFM}=(0.99\pm0.02) \times \theta_\mathrm{SPM}+
(0.02\pm0.07)$.

The surface-brightness method\cite{DiBenedetto98} (SBM) provides another way to
derive the angular diameter: 
\begin{equation} \label{eq:sb}
\frac{\theta_\mathrm{SBM} \, (\mathrm{mas})}{9.306\times10^{-0.2\,(V-A_V)}}
= 10^{0.2\,S_V}
\quad \mathrm{with} \quad
S_V = 2.536+1.493\,(V-K)_0-0.046(V-K)_0^2,
\end{equation}
where $S_V$ is the surface brightness in the V band. We have plotted on
Fig.~\ref{fig:comp}b the right-hand side of Eq.~\ref{eq:sb} vs. V$-$K. The
superimposed values for our reference stars (crosses) match the curve nicely.

These two results demonstrate that the spectro-photometric method is completely
consistent with other indirect methods.
%
%_____________________________________________ 5.2 Interferometric measurements
%
\subsection{Interferometric measurements}
The NPOI and Mark~III interferometers have measured the angular diameters of 21
stars belonging to our catalog\cite{Nordgren01}. We compare here the LD
diameters deduced from the UD diameters measured in the visible, using a
procedure very similar to the conversion process described in
Sect.~\ref{sub:diam}. Again, the agreement is very good (Fig.~\ref{fig:comp}a):
a linear least-square fit to the data yields
$\theta_\mathrm{LBSI}=(1.03\pm0.01) \times \theta_\mathrm{SPM}+(-0.15\pm0.03)$.
The average precisions of the NPOI and Mark~III data are respectively 1.9\% and
1.6\%. A chi-square analysis of the difference $\theta_\mathrm{SPM} -
\theta_\mathrm{LBSI}$ shows a good compatibility of the error bars since
$\chi^2$ equals respectively 3.0 and 2.4.
%
%_________________________________________________________________ 6 Conclusion
%
\section{CONCLUSION} \label{sec:concl}
We have presented a catalog of 374 carefully chosen reference stars for
optical interferometry. Depending on the needed precision on the visibility, it
is well suited for interferometers with baselines up to 200~m.
Although this catalog has proven to be fully satisfactory since its first use
by the FLUOR team in october 1999, most stars have not yet been observed by any
interferometer and still need to be checked. More work lies ahead to extend this
catalog to reference stars suitable for longer baselines, such as the baselines
of CHARA\cite{McAlister00} (330~m) and 'OHANA\cite{Perrin00} (800~m), or to
instruments with very high accuracies like AMBER\cite{Petrov00}.
%
%___________________________________________________________________ References
%

\end{document}